\documentstyle[12pt,qqa4lart]{article}
\input tcilatex

\QQQ{Language}{
American English
}

\begin{document}

\author{Lu-Ming Duan and Guan-Can Guo\thanks{%
Email: gcguo@sunlx06.nsc.ustc.edu.cn} \\
Department of Physics, University of Science \\
and Technology of China , Hefei, 230026, P.R.China}
\title{Cooperative loss and decoherence in quantum computation and commuication }
\date{}
\maketitle

\begin{abstract}
\baselineskip 24ptCooperative effects in the loss (the amplitude damping)
and decoherence (the phase damping) of the qubits (two-state quantum
systems) due to the inevitable coupling to the same environment are
investigated. It is found that the qubits undergo the dissipation coherently
in this case. In particular, for a special kind of input states (called the
coherence-preserving states), whose form depends on the type of the
coupling, loss and decoherence in quantum memory are much reduced. Based on
this phenomenon, a scheme by encoding the general input states of the qubits
into the corresponding coherence-preserving states is proposed for reducing
the cooperative loss and decoherence in quantum computation or communication.%
\\

\end{abstract}

\newpage\ \baselineskip 24pt

In recent years, quantum computation and communication have undergone a
dramatic evolution [1]. New algorithms such as factoring [2,3] were
developed and some individual quantum logic gates had been implemented in
experiments [4,5]. Quantum computers act as sophisticated, nonlinear
interferometers. The coherent interference pattern between the multitude of
superpositions is essential for taking advantage of quantum parallelism.
Unfortunately, decoherence of the qubits caused by the interaction with the
environment will collapse the state of the quantum computer and make the
result of the computation no longer correct [6]. To overcome this
difficulty, Shor [7] has shown it is possible to restore a desired state
using only partial knowledge of the state of the quantum computer. This
scheme is called quantum-error correction, which operates in a subtle way,
essentially by embedding the quantum information to be protected in a
subspace so oriented in a larger state space as to leak no or little
information to the environment. Many kinds of quantum-error correcting codes
have since been discovered which correct for specific interactions [8-15].

In the existing quantum-error correction codes, it is generally assumed that
the qubits decohere independently, which implies that the different qubits
couple to separate environments. A natural question is, what will occur if
the qubits couple cooperatively to the same environment? This situation may
be more practical. For example, the intrinsic decoherence in the ion trap
quantum computers just results from the cooperative coupling of the ions to
the environment [16]. If one only consider the phase damping, it has been
shown [17] that the cooperative dissipation results in the coherent
decoherence. And more interestingly, for a special kind of input states,
i.e., the coherence-preserving states, the qubits undergo no decoherence at
all even in the noisy memory. In this letter, we consider the general
dissipation, including the amplitude damping (loss) and the phase damping
(decoherence). The decoherence time is obtained for the qubits. Cooperative
effects in the loss and decoherence are examined. Interestingly, there still
exist the coherence-preserving states, whose form depends on the type of the
coupling between the qubits and the environment. For these states,
decoherence of the qubits is much recduced. Based on this phenomenon, we
furthermore propose a scheme for reducing the cooperative loss and
decoherence in quantum computation or communication. The scheme operates by
encoding the general input states of the qubits into the corresponding
coherence-preserving states in a slightly larger Hilbert space. The cost of
encoding $L$ qubits varies from $2L$ qubits to a function that approachs $L$
qubits asymptotically as L grows. So this scheme is very efficient.

The qubits in the memory, which may be spin-$\frac 12$ electrons or
two-level atoms, can be described by Pauli's operators $\overrightarrow{%
\sigma }_l$ ( $l$ marks different qubits ). The environment is modeled by a
bath of oscillators with infinite degrees of freedom. The general
dissipation of the qubits, including the phase damping and the amplitude
damping, is described by the following coupling Hamiltonian 
\begin{equation}
\label{1}H=\hbar \left\{ \omega _0\stackrel{L}{\stackunder{l=1}{\sum }}%
\sigma _l^z+\stackrel{L}{\stackunder{l=1}{\sum }}\stackunder{\omega }{\sum }%
\left[ \left( \lambda _{\omega l}^{\left( 1\right) }\sigma _l^x+\lambda
_{\omega l}^{\left( 2\right) }\sigma _l^y+\lambda _{\omega l}^{\left(
3\right) }\sigma _l^z\right) \left( a_\omega ^{+}+a_\omega \right) \right] +%
\stackunder{\omega }{\sum }\omega a_\omega ^{+}a_\omega \right\} ,
\end{equation}
where $L$ is the number of qubits. $a_\omega $ indicates the bath operator.
The coupling constants $\lambda ^{\left( 1\right) },\lambda ^{\left(
2\right) },\lambda ^{\left( 3\right) }$ may be independent of $\omega $ and $%
l$. This coupling system is nonlinear and very complicated, which makes it
impossible to find its exact solutions. However, in quantum computation or
communication, we mainly take interest in the decoherence time. This time
gives a mark after which the state of the qubits is obviously collapsed. To
obtain the decoherence time, we only need study the short-time behavior of
the coupling system. Very recently, Kim, etc., [18] propose a short-time
perturbative scheme for studying coherence loss. Here we follow this method.
The reduced density of the qubits is indicated by $\rho _1\left( t\right) $.
A simple and direct measure of the degree of decoherence of the qubits is
thus provided by the ''idempotency defect'' of the density $\rho _1\left(
t\right) $ [19], which is written in the equation below, where it is
furthermore subjected to a short-time power series expansion: 
\begin{equation}
\label{2}
\begin{array}{c}
\delta \left( t\right) =tr\left[ \rho _1\left( t\right) -\rho _1^2\left(
t\right) \right]  \\  
\\ 
=\delta _0+\frac t{\tau _1}+\frac{t^2}{\tau _2^2}+\cdots ,
\end{array}
\end{equation}
where $\delta _0,\tau _1,\tau _2$ can be expressed by $\rho \left( 0\right) $
(the initial density of the whole system), $H$ (the Hamiltonian) and their
commutators. Since the qubits and the environment are not entangled at the
beginning and usually the input states of the qubits are pure, $\rho \left(
0\right) $ can be factorized as $\rho \left( 0\right) =\left| \Psi _1\left(
0\right) \right\rangle \left\langle \Psi _1\left( 0\right) \right| \otimes
\rho _2\left( 0\right) $, where $\rho _2\left( 0\right) $ indicates the
reduced density of the environment, which is generally in a mixed state.
Under this condition, $\delta _0=\frac 1{\tau _1}=0$ and $\tau _2$, which
marks the decoherence time, is expressed as 
\begin{equation}
\label{3}\frac{\hbar ^2}{2\tau _2^2}=\left\langle H^2\right\rangle
_{1,2}+\left\langle H\right\rangle _{1,2}^2-\left\langle \left\langle
H\right\rangle _1^2\right\rangle _2-\left\langle \left\langle H\right\rangle
_2^2\right\rangle _1,
\end{equation}
where $\left\langle H\right\rangle _{1(2)}$ stands for the average of the
Hamiltonian over the subsystem $1(2)$, i.e., $\left\langle H\right\rangle
_1=\left\langle \Psi _1\left( 0\right) \right| H\left| \Psi _1\left(
0\right) \right\rangle $, $\left\langle H\right\rangle _2=tr\left( \rho
_2\left( 0\right) H\right) ,$ $\left\langle H\right\rangle
_{1,2}=\left\langle \Psi _1\left( 0\right) \right| \left\langle
H\right\rangle _2\left| \Psi _1\left( 0\right) \right\rangle $. Only with
pure input states for the qubits, the decoherence time can be simplified to
Eq.(3).

Now, consider the Hamiltonian (1), which can be rewritten as 
\begin{equation}
\label{4}H=H_1+H_2+\stackrel{L}{\stackunder{l=1}{\sum }}\stackrel{3}{%
\stackunder{\mu =1}{\sum }}H_{1l}^{\left( \mu \right) }H_{2l}^{\left( \mu
\right) },
\end{equation}
where $H_1=\hbar \omega _0\stackrel{L}{\stackunder{l=1}{\sum }}\sigma _l^z$, 
$H_2=\stackunder{\omega }{\sum }\hbar \omega a_\omega ^{+}a_\omega $, $%
H_{1l}^{\left( 1,2,3\right) }=\sigma _l^{\left( x,y,z\right) }$, $%
H_{2l}^{\left( \mu \right) }=\stackunder{\omega }{\sum }\hbar \lambda
_{\omega l}^{\left( \mu \right) }\left( a_\omega ^{+}+a_\omega \right) $.
The environment is supposed initially in thermal equilibrium, i.e., 
\begin{equation}
\label{5}\rho _2\left( 0\right) =\stackunder{\omega }{\prod }\int d^2\alpha
_\omega \frac 1{\pi \left\langle N_\omega \right\rangle }\exp \left( -\frac{%
\left| \alpha _\omega \right| ^2}{\pi \left\langle N_\omega \right\rangle }%
\right) \left| \alpha _\omega \right\rangle \left\langle \alpha _\omega
\right| 
\end{equation}
with the mean photon number 
\begin{equation}
\label{6}\left\langle N_\omega \right\rangle =1\left/ \left[ \exp \left( 
\frac{\hbar \omega }{k_BT}\right) -1\right] \right. .
\end{equation}
With the density (5) of the environment, we obviously have $\left\langle
H_{2l}^{\left( \mu \right) }\right\rangle =\left\langle H_2H_{2l}^{\left(
\mu \right) }\right\rangle =0$. Under this condition, Eq. (3) gives 
\begin{equation}
\label{7}\frac{\hbar ^2}{2\tau _2^2}=\stackunder{1\leq i,j\leq L}{\sum }%
\stackunder{\text{ }1\leq \mu ,\nu \leq 3}{\sum }\left\langle H_{2i}^{\left(
\mu \right) }H_{2j}^{\left( \nu \right) }\right\rangle \left( \left\langle
H_{1i}^{\left( \mu \right) }H_{1j}^{\left( \nu \right) }\right\rangle
-\left\langle H_{1i}^{\left( \mu \right) }\right\rangle \left\langle
H_{1j}^{\left( \nu \right) }\right\rangle \right) .
\end{equation}
In practice, the coupling constants $\lambda _{\omega l}^{\left( \mu \right)
}$ $\left( \mu =1,2,3\right) $ often factor as $\lambda _{\omega l}^{\left(
\mu \right) }=\lambda _l^{\left( \mu \right) }\kappa \left( \omega \right) $%
. Then Eq. (7) can be further simplified. To show this, let 
\begin{equation}
\label{8}\Omega ^2=2\left\langle \left[ \stackunder{\omega }{\sum }\kappa
\left( \omega \right) \left( a_\omega ^{+}+a_\omega \right) \right]
^2\right\rangle =4\int d\omega \kappa ^2\left( \omega \right) \left(
\left\langle N_\omega \right\rangle +\frac 12\right) 
\end{equation}
and define 
\begin{equation}
\label{9}A=\stackunder{l,\mu }{\sum }\lambda _l^{\left( \mu \right)
}H_{1l}^{\left( \mu \right) }=\stackrel{L}{\stackunder{l=1}{\sum }}\left(
\lambda _l^{\left( 1\right) }\sigma _l^x+\lambda _l^{\left( 2\right) }\sigma
_l^y+\lambda _l^{\left( 3\right) }\sigma _l^z\right) ,
\end{equation}
the decoherence time $\tau _2$ thus becomes 
\begin{equation}
\label{10}\frac 1{\tau _2^2}=\Omega ^2\left\langle \left( \Delta A\right)
^2\right\rangle .
\end{equation}

Eq.(10) suggests that the qubits decohere coherently when they couple
cooperatively to the same environment. This fact can be more clearly seen by
comparison with the decoherence in the case that the qubits interact
independently with separate environments. Very simple calculation yields the
decoherence time in the latter situation 
\begin{equation}
\label{11}\frac 1{\tau _2^{^{\prime }2}}=\Omega ^2\stackrel{L}{\stackunder{%
l=1}{\sum }}\left\langle \left( \Delta A_l\right) ^2\right\rangle , 
\end{equation}
where $A_l=\lambda _l^{\left( 1\right) }\sigma _l^x+\lambda _l^{\left(
2\right) }\sigma _l^y+\lambda _l^{\left( 3\right) }\sigma _l^z$. The
decoherence rate $\frac 1{\tau _2^{^{\prime }2}}$ increases with $L$
monotonically. Its typical behavior is $\frac 1{\tau _2^{^{\prime
}2}}\propto L$. This decoherence is insensitive to the input states of the
qubits. In contrast, for the cooperative dissipation of the qubits, the
decoherence rate $\frac 1{\tau _2^2}$ depends greatly on the type of the
input states. For some input states, $\frac 1{\tau _2^2}$ increases with $L$
very rapidly, whereas for some other input states, the decoherence rate does
not increase with $L$ at all. In particular, for the eigenstates of the
operator $A$, $\frac 1{\tau _2^2}=0$, which suggests, for this kind of input
states the decoherence is much reduced.

We briefly discuss the eigenstates of the operator $A$, which may be called
the coherence-preserving states. The case that the coupling constants $%
\lambda _l^{\left( \mu \right) }$ are equal for different qubits $l$ is of
special interest. Then $\lambda _l^{\left( \mu \right) }$ is just indicated
by $\lambda ^{\left( \mu \right) }$. The Hermitian operators $A_l$ satisfy $%
tr\left( A_l\right) =0$, so their eigenvalues are $\pm a$, where $a$ is a
real number. Without loss of generality, their corresponding eigenstates can
be indicated by $\left| \pm 1\right\rangle _l$. For example, if there is
only the phase damping, $\lambda ^{\left( 1\right) }=\lambda ^{\left(
2\right) }=0$, so $\left| \pm 1\right\rangle _l=\left| \pm \right\rangle _l$%
, where $\left| \pm \right\rangle _l$ stand for the eigenvectors of $\sigma
_l^z$. On the other hand, if there is only the amplitude damping, i.e., $%
\lambda ^{\left( 2\right) }=\lambda ^{\left( 3\right) }=0,$ $\left| \pm
1\right\rangle _l=\frac 1{\sqrt{2}}\left( \left| +\right\rangle _l\pm \left|
-\right\rangle _l\right) $, which are the eigenvectors of $\sigma _l^x$. The
eigenstates of the operator $A$ can be easily constructed from the states $%
\left| \pm 1\right\rangle _l$. They are 
\begin{equation}
\label{12}\left| \Psi _{2L}\right\rangle _m=\stackunder{\left\{
i_1,i_2,\cdots ,i_{2L}\left| \stackrel{2L}{\stackunder{l=1}{\sum }}%
i_l=m\right. \right\} }{\sum }c_{\left\{ i_l\right\} }\left| \left\{
i_l\right\} \right\rangle ,
\end{equation}
where the constants $m$ stand for the eigenvalues of the operator $A$. The
state $\left| \left\{ i_l\right\} \right\rangle $ indicates $\left|
i_1\right\rangle \otimes \left| i_2\right\rangle \otimes \cdots \otimes
\left| i_{2L}\right\rangle $ and $i_l=\pm 1$. Here we have supposed there
are $2L$ qubits. With $m=0,\pm 2,\pm 4,\cdots ,\pm 2L$, the dimensions of
the state spaces expanded by the vectors $\left| \Psi _{2L}\right\rangle
_0,\left| \Psi _{2L}\right\rangle _{\pm 2},\cdots ,\left| \Psi
_{2L}\right\rangle _{\pm 2L}$ are respectively $\left( 
\begin{array}{c}
2L \\ 
L
\end{array}
\right) ,2\left( 
\begin{array}{c}
2L \\ 
L-1
\end{array}
\right) ,\cdots ,2\left( 
\begin{array}{c}
2L \\ 
0
\end{array}
\right) $. The sum of these dimensions satisfies%
$$
\left( 
\begin{array}{c}
2L \\ 
L
\end{array}
\right) +2\left( 
\begin{array}{c}
2L \\ 
L-1
\end{array}
\right) +\cdots +2\left( 
\begin{array}{c}
2L \\ 
0
\end{array}
\right) =2^{2L}. 
$$
So all the eigenstates of the operator $A$ make a complete basis for the
Hilbert space of $2L$ qubits.

Now the question is how to exploit these coherence-preserving states to
reduce the cooperative decoherence in quantum memory. In the following we
show this may be achieved by encoding arbitrary input states of the qubits
into the corresponding coherence-preserving states in a larger state space.
The input states of $L$ qubits can be generally expressed as 
\begin{equation}
\label{13}\left| \Psi _L\right\rangle =\stackunder{\left\{ i_l\right\} }{%
\sum }c_{\left\{ i_l\right\} }\left| \left\{ i_l\right\} \right\rangle , 
\end{equation}
where $\left| \left\{ i_l\right\} \right\rangle $ is just the abbreviation
of $\left| i_1,i_2,\cdots ,i_L\right\rangle $. If only one qubit with the
state $c_1\left| 1\right\rangle +c_{-1}\left| -1\right\rangle $ is input, we
encode the input state into the state $c_1\left| 1,-1\right\rangle
+c_{-1}\left| -1,1\right\rangle $ of two qubits. Obviously, the latter is a
coherence-preserving state. Similarly, there is one-to-one correspondence
between the input states of $L$ qubits and the following
coherence-preserving states in the Hilbert space of $2L$ qubits 
\begin{equation}
\label{14}\left| \Psi _{2L}\right\rangle _{coh}=\stackunder{\left\{
i_l\right\} }{\sum }c_{\left\{ i_l\right\} }\left| \left\{ i_l,-i_l\right\}
\right\rangle , 
\end{equation}
where $\left| \left\{ i_l,-i_l\right\} \right\rangle $ represents $\left|
i_1,-i_1,i_2,-i_2,\cdots ,i_L,-i_L\right\rangle $. We encode the input
states (13) into the corresponding states in the form of Eq. (14) before
storing them into the memory. The encoded states undergo reduced decoherence
in the noisy memory and then they can be decoded into the original states.
By this scheme the cooperative decoherence is much reduced, especially when
the storing time is short.

The encoding and decoding in the above scheme can be easily realized in
quantum computers by the elementary logic gates, the quantum controlled-NOT.
Some applications of this sort of logic gates have been commented in Ref.
[20] with stess on the appearance of a conditional quantum dynamics. The
quantum controlled-NOT gate is defined as that which effects the unitary
operation on two qubits, which in a chosen orthonormal basis $\left\{ \left|
-1\right\rangle ,\left| 1\right\rangle \right\} $ reproduces the classical
controlled-NOT operation 
\begin{equation}
\label{15}\left| \varepsilon _1\right\rangle _1\left| \varepsilon
_2\right\rangle _2\stackrel{C_{12}}{\longrightarrow }\left| \varepsilon
_1\right\rangle _1\left| -\varepsilon _1\cdot \varepsilon _2\right\rangle _2,
\end{equation}
Here and in the following the first subscript of $C_{ij}$ refers to the
control bit and the second to the target bit. To encode $L$ qubits in the
state (13), $L$ ancillary qubits $1^{^{\prime }},2^{^{\prime }},\cdots
,L^{^{\prime }}$ need be prearranged in the state $\left| \Psi _{1^{^{\prime
}}2^{^{\prime }}\cdots L^{^{\prime }}}\right\rangle =\left| 1\right\rangle
_{1^{^{\prime }}}\otimes \left| 1\right\rangle _{2^{^{\prime }}}\otimes
\cdots \otimes \left| 1\right\rangle _{L^{^{\prime }}}$. From the definition
(15), the state (13) is transformed into the coherence-preserving state (14)
of the $2L$ qubits $1,1^{^{\prime }},2,2^{^{\prime }},\cdots ,L,L^{^{\prime
}}$ by $L$ times controlled-NOT operations 
\begin{equation}
\label{16}\left| \Psi _{12\cdots L}\right\rangle \otimes \left| \Psi
_{1^{^{\prime }}2^{^{\prime }}\cdots L^{^{\prime }}}\right\rangle \stackrel{%
C_{11^{^{\prime }}}C_{22^{^{\prime }}}\cdots C_{LL^{^{\prime }}}}{%
\longrightarrow \longrightarrow }\stackunder{\left\{ i_l\right\} }{\sum }%
c_{\left\{ i_l\right\} }\left| \left\{ i_l,-i_l\right\} \right\rangle
=\left| \Psi _{11^{^{\prime }}22^{^{\prime }}\cdots LL^{^{\prime
}}}\right\rangle _{coh}.
\end{equation}
This transformation can be reversed by applying the same controlled-NOT
operations again. So the decoding is fulfilled by
\begin{equation}
\label{17}\left| \Psi _{11^{^{\prime }}22^{^{\prime }}\cdots LL^{^{\prime
}}}\right\rangle _{coh}\stackrel{C_{11^{^{\prime }}}C_{22^{^{\prime
}}}\cdots C_{LL^{^{\prime }}}}{\longrightarrow \longrightarrow }\left| \Psi
_{12\cdots L}\right\rangle \otimes \left| \Psi _{1^{^{\prime }}2^{^{\prime
}}\cdots L^{^{\prime }}}\right\rangle .
\end{equation}

The above encoding scheme is very simple but not efficient, since only the
coherence-preserving states in the form of Eq. (14) are used. Suppose there
are $2L$ qubits. The maximum dimension of the eigenspace of the operator $A=%
\stackrel{2L}{\stackunder{l=1}{\sum }}A_l$ is $\left( 
\begin{array}{c}
2L \\ 
L
\end{array}
\right) $, with the eigenvalue $m=0$. If all the coherence-preserving states
in this eigenspace are fully used, the efficiency $\eta $ of the encoding
attains the maximum, which is 
\begin{equation}
\label{18}\eta _M=\log _2\frac{\left( 
\begin{array}{c}
2L \\ 
L
\end{array}
\right) }{2^{2L}}\approx 1-\frac 1{4L}\log _2\left( \pi L\right) .
\end{equation}
In Eq. (18) the approximation $L>>1$ is introduced and the stirling formula $%
L!\approx \sqrt{2\pi }L^{L+\frac 12}e^{-L}$ is used. As $L$ grows, the
efficiency $\eta _M$ tends to $1$. So in the perfect encoding scheme, the
input states can be transformed into the coherence-preserving states almost
without expanding of the number of qubits. Of course, to raise the
efficiency, the encoding scheme will correspondingly become complicated and
involve much more logic gates.

It is interesting to compare this strategy with the quantum-error correction
schemes. In the error-correction schemes, the decoherence time for a qubit
is not increased. What one does is to repeatedly restore the original state
of the qubits from the decohered encoded state by unitary transformations
and measurements on some ancillary qubits. The quantum state should be
restored at time intervals much less than the decoherence time, and to make
the restoration possible, one needs at least $5L$ qubits to encode $L$
qubits [12,15]. On the other hand, in the present scheme what we do is to
increase the decoherence time. We deal with the case that the qubits
decohere cooperatively. The cooperative dissipation of the qubits results in
the coherence-preserving states. By encoding the input states into the
corresponding coherence-preserving states, the decoherence time for a qubit
is increased. The cost of encoding $L$ qubits varies from $2L$ qubits to $%
L+\frac 12\log _2\left( \frac \pi 2L\right) $ qubits (from Eq. (18)). So by
this scheme the decoherence is reduced at little cost.

In this letter, the decoherence time is obtained by short-time expansion. If
the coherence-preserving states are input, the second-order decoherence rate
equals zero. What about the higher order contributions? It has been shown
that if there is only the phase damping, all the higher order contributions
disappear at the same time [17]. So in this case the cooperative dcoherence
can be eliminated by encoding the input states into the coherence-preserving
states. But in general cases, especially when the amplitude damping is
dominant, the higher order contributions do not equal zero. Therefore, by
this scheme the decoherence is reduced but can not be eliminated. This is
the main disadvantage of the scheme. However, this shortcoming can be
overcome by combining the scheme with the quantum-error correction. In
quantum-error correction, if the unitary transformations and the
measurements are perfect, the error rate can be made arbitrarily small by
repeatedly restoring the quantum state. But in practice, each time one gets
rid of the decoherence, he introduces some extra error. So there is also a
small amount of error which is hard to be eliminated by the error correction
schemes. However, if we combine these two schemes together, it is possible
to further reduce the error rate, and also, the efficiency of the encoding
may be raised. So it is of interest to find a scheme for correcting
quantun-error caused by the cooperative decoherence. This question needs
further investigation, since the existing quantum-error correction schemes
are devoted to reducing the independent decoherence.

\section*{Acknowledgment}

This project was supported by the National Nature Foundation of China.

\newpage\

\end{document}